\documentclass[aps,prb,preprint,superscriptaddress,colorlinks=true,linkcolor=black,urlcolor=black,citecolor=black,anchorcolor=black,colorlinks,bookmarks=false,citecolor=blue,linkcolor=black,urlcolor=blue]{revtex4-1}

\usepackage{ulem}
\usepackage{bm,color,amsmath,amssymb,latexsym,graphicx,psfrag,accents,float}
\usepackage{amscd,dsfont,wasysym,mathrsfs,mathtools}
\usepackage{multirow}
\usepackage{dcolumn}
\usepackage{xcolor}
\usepackage{comment}
\usepackage{booktabs} 
\usepackage{upgreek}
\usepackage[utf8]{inputenc}
\usepackage{braket}
\usepackage{times}
\usepackage{amsfonts}
\usepackage{mathrsfs}
\usepackage{graphicx}
\usepackage{dcolumn}
\usepackage{bm}
\usepackage{color}

\usepackage[colorlinks,bookmarks=false,citecolor=blue,linkcolor=red,urlcolor=blue]{hyperref}
\usepackage{amsmath}
\usepackage{ulem}
\begin{document}
\title{3D Fermi surfaces from charge order in layered CsV$_3$Sb$_5$}
\author{Xiangwei Huang${}^{}$}\affiliation{Laboratory of Quantum Materials (QMAT), Institute of Materials (IMX),\'{E}cole Polytechnique F\'{e}d\'{e}rale de Lausanne (EPFL), CH-1015 Lausanne, Switzerland}
\author{Chunyu Guo${}^{\ast}$}\affiliation{Laboratory of Quantum Materials (QMAT), Institute of Materials (IMX),\'{E}cole Polytechnique F\'{e}d\'{e}rale de Lausanne (EPFL), CH-1015 Lausanne, Switzerland}
\affiliation{Max Planck Institute for the Structure and Dynamics of Matter, 22761 Hamburg, Germany}
\author{Carsten Putzke${}^{}$}\affiliation{Laboratory of Quantum Materials (QMAT), Institute of Materials (IMX),\'{E}cole Polytechnique F\'{e}d\'{e}rale de Lausanne (EPFL), CH-1015 Lausanne, Switzerland}
\affiliation{Max Planck Institute for the Structure and Dynamics of Matter, 22761 Hamburg, Germany}
\author{Martin Gutierrez-Amigo}
\affiliation{Department of Physics, University of the Basque Country (UPV/EHU), 48080 Bilbao, Spain}
\affiliation{Centro de Física de Materiales (CSIC-UPV/EHU), 20018 Donostia-San Sebastian, Spain}

\author{Yan Sun}\affiliation{Max Planck Institute for Chemical Physics of Solids, 01187 Dresden, Germany}
\author{Maia G. Vergniory}\affiliation{Max Planck Institute for Chemical Physics of Solids, 01187 Dresden, Germany}
\affiliation{Donostia International Physics Center, 20018 Donostia-San Sebastian, Spain}

\author{Ion Errea}\affiliation{Donostia International Physics Center, 20018 Donostia-San Sebastian, Spain}
\affiliation{Centro de Física de Materiales (CSIC-UPV/EHU), 20018 Donostia-San Sebastian, Spain}
\affiliation{Fisika Aplikatua Saila, Gipuzkoako Ingeniaritza Eskola, University of the Basque Country (UPV/EHU), 20018 Donostia-San Sebastian, Spain}

\author{Dong Chen}\affiliation{Max Planck Institute for Chemical Physics of Solids, 01187 Dresden, Germany}

\author{Claudia Felser}\affiliation{Max Planck Institute for Chemical Physics of Solids, 01187 Dresden, Germany}
\author{Philip J. W. Moll${}^{\dagger}$}\affiliation{Laboratory of Quantum Materials (QMAT), Institute of Materials (IMX),\'{E}cole Polytechnique F\'{e}d\'{e}rale de Lausanne (EPFL), CH-1015 Lausanne, Switzerland}
\affiliation{Max Planck Institute for the Structure and Dynamics of Matter, 22761 Hamburg, Germany}
\date{\today}

\maketitle

\section*{Abstract}
The cascade of electronic phases in CsV$_3$Sb$_5$ raises the prospect to disentangle their mutual interactions in a clean, strongly interacting Kagome lattice. When the Kagome planes are stacked into a crystal, its electronic dimensionality encodes how much of the Kagome physics and its topological aspects survive. The layered structure of CsV$_3$Sb$_5$ reflects in Brillouin-zone-sized quasi-2D Fermi surfaces and a significant transport anisotropy. Yet here we demonstrate that CsV$_3$Sb$_5$ is a three-dimensional metal within the charge-density-wave (CDW) state. Small 3D pockets play a crucial role in its low-temperature magneto- and quantum transport. Their emergence at $T_{CDW}\sim 93 K$ results in an anomalous sudden increase of the in-plane magnetoresistance by 4 orders of magnitude. The presence of these 3D pockets is further confirmed by quantum oscillations under in-plane magnetic fields - demonstrating their closed nature. These results emphasize the impact of interlayer coupling on the Kagome physics in 3D materials.

\section*{introduction}

When structurally layered materials host strong electronic correlations, their effective electronic dimensionality is key to understanding their microscopic physics. The anisotropic Fermi surface then sets the canvas on which correlated ground states unfold, such as magnetism, charge-density-waves (CDW) or superconductivity. Cuprate and pnictide high-temperature superconductors are prime examples of layered materials in which reduced dimensionality defines the superconducting state\cite{bednorz1986possible,kamihara2006iron}. The anisotropy is further a critical parameter to describe the vortex formation and the orbital limit of the upper critical field. Recently, metals hosting planes with Kagome nets have attracted significant attention due to their topological phases as well as potential correlation effects\cite{kida2011giant,kang2020topological,ortiz2019new}. The latter have been argued to arise from small-bandwidth bands (flat bands) that are predicted in phenotypical two-dimensional Kagome models\cite{o2010strongly}. Flat bands naturally enhance correlation effects when the Coulomb energy exceeds the kinetic energy and they lead to a divergent density of states, an entropic catastrophe which is commonly avoided by the formation of electronically ordered states at low temperatures. However when the Kagome nets are stacked into a three-dimensional crystal, the existence of flat bands depends on the strength of the interlayer hybridization and its impact on the band dispersion. Therefore in order to explore how much Kagome physics survives in a 3D metal, it is necessary to investigate the electronic structure and especially its effective dimensionality.  

Here we investigate the electronic dimensionality of CsV$_{3}$Sb$_{5}$ [Fig. 1(a) and (b)], a stacked Kagome crystal, and its influence on the rich physics it hosts such as the  non-trivial band topology, superconductivity and CDW order. The non-magnetic Kagome nets formed by Vanadium atoms result in a symmetry-enforced pair of $\mathbb{Z}_2$ topological bands with opposite Chern numbers \cite{ortiz2019new,ortiz2020cs,fu2021quantum,CDANE}. The hybridization with Sb atoms effectively expands the band width and therefore the flat band gains sizable dispersion yet it leaves multiple bulk Dirac points intact, as observed by angle-resolved photoemission spectroscopy (ARPES)\cite{ortiz2020cs,luo2021distinct}. Upon cooling, the Kagome lattice undergoes a CDW transition at $T_{CDW}$ = 93 K \cite{yu2021concurrence,mu2021s,liang2021three}. The 2$\times$2 reconstruction within the Kagome plane is accompanied by a $\pi$ phase shift of the CDW across an atomic step edge as observed by STM experiments \cite{zhao2021cascade,liang2021three}, suggesting a band reconstruction along the $c$-direction. X-ray studies further suggest a 2$\times$2$\times$4 superlattice formation \cite{Xray,XrayCD} with long range out-of-plane coherence, again highlighting the importance of out-of-plane coupling in this system. At even lower temperatures of $T_{c} \sim 2.8$ K, superconductivity appears\cite{ortiz2020cs}. The large upper critical field anisotropy ($\sim$~9) is a result of its anisotropic superconducting properties\cite{ni2021anisotropic}. Meanwhile the non-trivial band topology may give rise to topological superconductivity. Indeed, STM has uncovered a zero-bias peak in the vortex cores which, while not conclusive, is compatible with a scenario of bound Majorana states\cite{liang2021three}.


\section*{Results}


The single-particle band structure from ab-initio calculations serves as a starting point to investigate the dimensionality of this strongly correlated compound. The electronic structure of CsV$_3$Sb$_5$ in the high-symmetry state at room temperature is calculated using density functional theory (DFT) methods, the computational details can be found in the Supplementary materials\cite{QE-2017,perdew1996generalized,DALCORSO2014337,MP-smearing,mostofi_2014}. Multiple bands are found at the Fermi level [Fig. 1(c)], consistent with the previous reports\cite{ortiz2019new,ortiz2020cs,fu2021quantum}. Accordingly, multiple Fermi surfaces of cylindrical shape are expected in the Brillouin zone [Fig. 1(d)]. While the layered structure is reflected in the electron-like cylinders centered at $\Gamma$ and K, their sizable dispersion signals pronounced interlayer coupling. In absence of spin-orbit coupling (SOC), the cylinder at the Brillouin zone boundary (H to K) would surround a nodal line, yet SOC breaks this degeneracy gapping the nodal line.


The band structure of CsV$_3$Sb$_5$ indicates a clear out-of-plane dispersion, giving rise to warping of the cylindrical Fermi surfaces. Experimentally the electronic dimensionality can be explored via the resistivity anisotropy. As common in structurally layered materials,  CsV$_3$Sb$_5$ crystals grow as thin platelets along the Kagome plane \cite{ortiz2019new}. While this crystal morphology lends itself to in-plane resistivity measurements, quantitative out-of-plane transport poses a well-known challenge. Focused Ion Beam (FIB) milling can be used to prepare micron-sized $c$-direction bars with well-defined geometries to quantify the resistivity\cite{moll2018focused}. CsV$_3$Sb$_5$ like other soft CDW-compounds is susceptible to mechanical strain\cite{song2021competing,du2021pressure}, we suspend the microstructured sample (purple) in free space. It is mechanically and electrically connected to a supportive silicon frame only via thin, gold-coated SiN$_x$ membrane microsprings (gold)  [Fig. 1(e) and (f)]. These structures exert only minimal residual pressure on the sample ($\approx$ 9.8~bar, see supplement), thus allowing an intrinsic evolution of the ordered states. Moreover, a low-voltage polish step at 5~kV reduces the amorphization layer into the nm-range\cite{kelley2013xe+} which minimizes its possible influence.



In these suspended structures, the in-plane resistivity decreases from 80~$\mu \Omega$cm at 300~K to 0.3~$\mu \Omega$cm at $T_c=2.8$~K. The high residual resistivity ratio (RRR) of 250 is comparable to bulk samples\cite{CDANE}, evidencing the high quality of the microstructure fabrication (Fig. 2). The in-plane resistivity displays a small yet sharply defined discontinuity at the CDW transition at $T_{CDW}$ which is consistent with previous reports\cite{CDANE,ortiz2019new,xiang2021twofold}. The out-of-plane resistivity, $\rho_c$, is comparably larger yet shows an overall similar metallic temperature dependence. It decreases from $\rho_c$(300K)~$\sim$ 1.9~m$\Omega$cm to $\rho_c$($T_c$)~$\sim$ 33~$\mu \Omega$cm, corresponding to a lower RRR of 60. These results are qualitatively reflected in the challenging measurements of $\rho_c$ on bulk crystals~\cite{ortiz2020cs,xiang2021twofold}, yet deviate quantitatively. In comparison to the in-plane resistivity, $\rho_c$ shows an upwards jump at $T_{CDW}$ [Fig. 2(c)]. This jump $(\rho_c(T_{CDW}^{+})-\rho_c(T_{CDW}^{-}))/\rho_c(T_{CDW}^{+}) \approx 12\%$ is much more pronounced compared to the change of $\approx -3\%$ for in-plane resistivity. This relative difference again emphasizes the importance of the CDW in the out-of-plane direction.


Further information about the electronic dimensionality is contained in the temperature dependence of the anisotropy. At high temperatures above $T_{CDW}$, the picture of an anisotropic metal emerges [$\rho_c/\rho_{ab} (T=300~K) \approx 24$], which falls close to the Fermi velocity anisotropy when averaged over the entire Fermi surface ($ |v_{F,\parallel}|/|v_{F,\perp}| \sim 20 $, [Fig. 2(d)]). The dominant carrier density originates from the Brillouin-zone-sized hexagonal Fermi surfaces with weak $k_z$ dispersion. These feature six almost flat surfaces parallel to the crystalline $a$-directions [Fig. 1(b)], which play an important role in the magneto-transport as discussed later.



Further lowering the temperature beyond the jump at $T_{CDW}$ increases the anisotropy which eventually saturates around 100 at $T_c$, suggesting dominant in-plane electronic transport at low temperatures. Yet $\rho_c$ reaches a low value of 33~$\mu \Omega cm$ at $T_c$, which signals metallic three-dimensional transport. The first-order derivative of the resistivity anisotropy reveals further, more subtle changes in the material [Fig. 2(e)]. Besides the clear spike at $T_{CDW}$, a local minimum occurs at around 70 K. This temperature coincides with the onset of anisotropy in the muon spin depolarization rate that has been associated with time-reversal-symmetry breaking (TRSB)\cite{yu2021evidence}. The TRSB state has been theoretically proposed to arise from an effective orbital current loop flux\cite{TitusAdd,feng2021chiral}, and a modification of a magnetic scattering channel would be a natural connection between this experiment and ours.

This further emphasizes the emerging question about its effective electronic dimensionality. The large resistivity anisotropy is compatible with an effective 2D description, which may be captured by a simplified model based on a 2D Kagome lattice. Yet to discuss the effective dimensionality, one has to carefully distinguish between the transport dimensionality and the dimensionality of the Fermi surface. In general, the transport anisotropy is determined by the Fermi velocity $v_F(k)$ and the scattering time $\tau (k)$ distribution while the Fermi surface topology encodes another aspect of electronic dimensionality. For example, a cylindrical Fermi surface with strong warping naturally features low transport anisotropy. Here we demonstrate that CsV$_3$Sb$_5$ is a 3D Kagome metal as the emergent small, closed Fermi pockets play a crucial role in its magneto- and quantum transport properties.

The magnetoresistance provides further information about the electronic dimensionality (Fig. 3). Most intriguing is the complete absence of transverse magnetoresistance for orthogonal in-plane fields and currents at any temperature above $T_{CDW}$. The field-independent noise floor provides an upper bound for the magnetoresistance, at{ ($\rho_a$(18 T)-$\rho_a$(0 T))/$\rho_a$(0 T) $<$ 10$^{-5}$} at 120~K. Even rather unsuspicious metals such as Cu commonly show transverse magnetoresistance around $1\%$~(20T) at room temperature and $100\%$~(20T) at 100K\cite{de1959magnetoresistance}. In contrast, for all other orthogonal current and field orientations in CsV$_3$Sb$_5$, the magnetoresistance increases with decreasing temperature following a conventional semiclassical scaling of the magneto-transport by $\omega_c \tau$. Given the complex Fermi surface and putative incipient TRSB order, such absent magnetoresistance for one particular direction appears quite exotic. 


The reconstruction of the main cylindrical Fermi surfaces at $T_{CDW}$ offers a natural explanation for such behavior. When a current is applied in the plane, the conductivity is dominated by the weakly warped main Fermi surfaces [Fig. 3(d)]. Such flatness ensures that the quasiparticle velocity remains approximately unchanged by the Lorentz force. This behavior is commonly observed in quasi-2D materials when only cylindrical Fermi surfaces are present\cite{TMTSF1,wosnitza2006fermi}. For magnetic field or current applied along any other direction, magnetoresistance is bound to appear as the Lorentz force significantly impacts the carrier trajectories. These results consistently demonstrate that  CsV$_3$Sb$_5$ is electronically quasi-2D above $T_{CDW}$.


However, this conclusion no longer holds in the charge ordered phase. Though the exact type of CDW reconstruction is still under debate, all different scenarios predict the emergence of small closed 3D pockets as a direct consequence of zone folding\cite{ortiz2021fermi}. Compared to the quasi-2D Fermi surfaces, these small pockets support closed cyclotron orbits which leads to the sudden recovery of magnetoresistance [Fig. 3(d)]. The direct connection between the abrupt change of magnetoresistance and Fermi surface reconstruction further demonstrates the importance of the small, closed pockets to the 3D electronic transport properties of CsV$_3$Sb$_5$, especially at low temperatures.


Such small, 3D-like pockets at low temperature should be visible in quantum oscillations. Four low frequency quantum oscillations have been reported for out-of-plane fields and associated with these small 3D pockets\cite{ortiz2021fermi,yu2021concurrence}. However, these quantum oscillations were observed to quickly disappear as the field is rotated towards the plane, which precluded the distinction between a closed, 3D pocket and an open, 2D cylinder. The weakly perturbed, suspended microbars now allow to track this frequency over the entire angle range, firmly establishing them as closed 3D pockets within the density wave phase (Fig. 4). At low angles, the oscillation frequencies observed in our structures are consistent with these previous reports. A likely scenario for the loss of amplitude in conventional experiments involves stacking disorder along $c$-direction due to residual strain, which reduces the quantum-coherent transport along the out-of-plane direction. This is consistent with the larger RRR and lower resistivity observed in the microstructures compared to macroscopic crystals\cite{ortiz2020cs,xiang2021twofold}. The angular dependence of the oscillation frequencies is well described by a simple ellipsoidal model $F_i \propto 1 /\sqrt{sin^2(\theta)+(1+A_i^2)cos^2(\theta)}$. The factor $A_i$ characterizes the anisotropy of the i-th Fermi surface [Fig. 4(d)]. These results further support the presence of 3D pockets as suggested by the unusual magneto-transport. The emergence of 3D Fermi pockets due to CDW reconstruction has already been observed in other CDW materials \cite{TiSe2-PRL,Tam2022} as a 3D zone-folding of Brillouin-zone-sized Fermi surface naturally results in small Fermi surface pockets. These pockets are connected by the 3/4 Bragg wavevector in the 3Q directions which results in an exotic pairing density wave phase at low temperature\cite{Theonew}, as observed in STM experiments \cite{Chen2021}.

In conclusion, the electronic anisotropy and Fermiology of charge-ordered CsV$_3$Sb$_5$ are characterized by the coexistence of a high-mobility in-plane electronic system with small 3D pockets. They emerge when the charge order is established, demonstrating an electronic dimensionality crossover at $T_{CDW}$. Therefore the interlayer coupling must be taken into account for theoretical modelling as considerations based on 2D Kagome lattices may miss important aspects of this material. It will be highly interesting to explore how the cascades of correlated electronic states are influenced by the coupling between Kagome nets.

\clearpage

\section*{References}

\clearpage

\noindent \textbf{Acknowledgements}
\\

\noindent \textbf{Funding: } This work was funded by the European Research Council (ERC) under the European Union’s Horizon 2020 research and innovation programme (MiTopMat - grant agreement No. 715730). This project received funding by the Swiss National Science Foundation (Grants  No. PP00P2\_176789). M.G.V., I. E. and M.G.A. acknowledge the Spanish   Ministerio de Ciencia e Innovacion  (grant PID2019-109905GB-C21). M.G.V. thanks support to Programa Red Guipuzcoana de Ciencia Tecnolog\'{i}a e Innovaci\'{o}n 2021 No. 2021-CIEN-000070-01 Gipuzkoa Next and the Deutsche Forschungsgemeinschaft (DFG, German Research Foundation) GA 3314/1-1 – FOR 5249 (QUAST). This work has been supported in part by Basque Government grant IT979-16. This work was also supported by the European Research Council Advanced Grant (No. 742068) “TOPMAT”, the Deutsche Forschungsgemeinschaft (Project-ID No. 258499086) “SFB 1143”, and the DFG through the W\"{u}rzburg-Dresden Cluster of Excellence on Complexity and Topology in Quantum Matter ct.qmat (EXC 2147, Project-ID No. 39085490).

\noindent \textbf{Competing Interests} The authors declare that they have no competing financial interests.
\clearpage
\begin{figure}
	\centering
	\includegraphics[width = 0.9\linewidth]{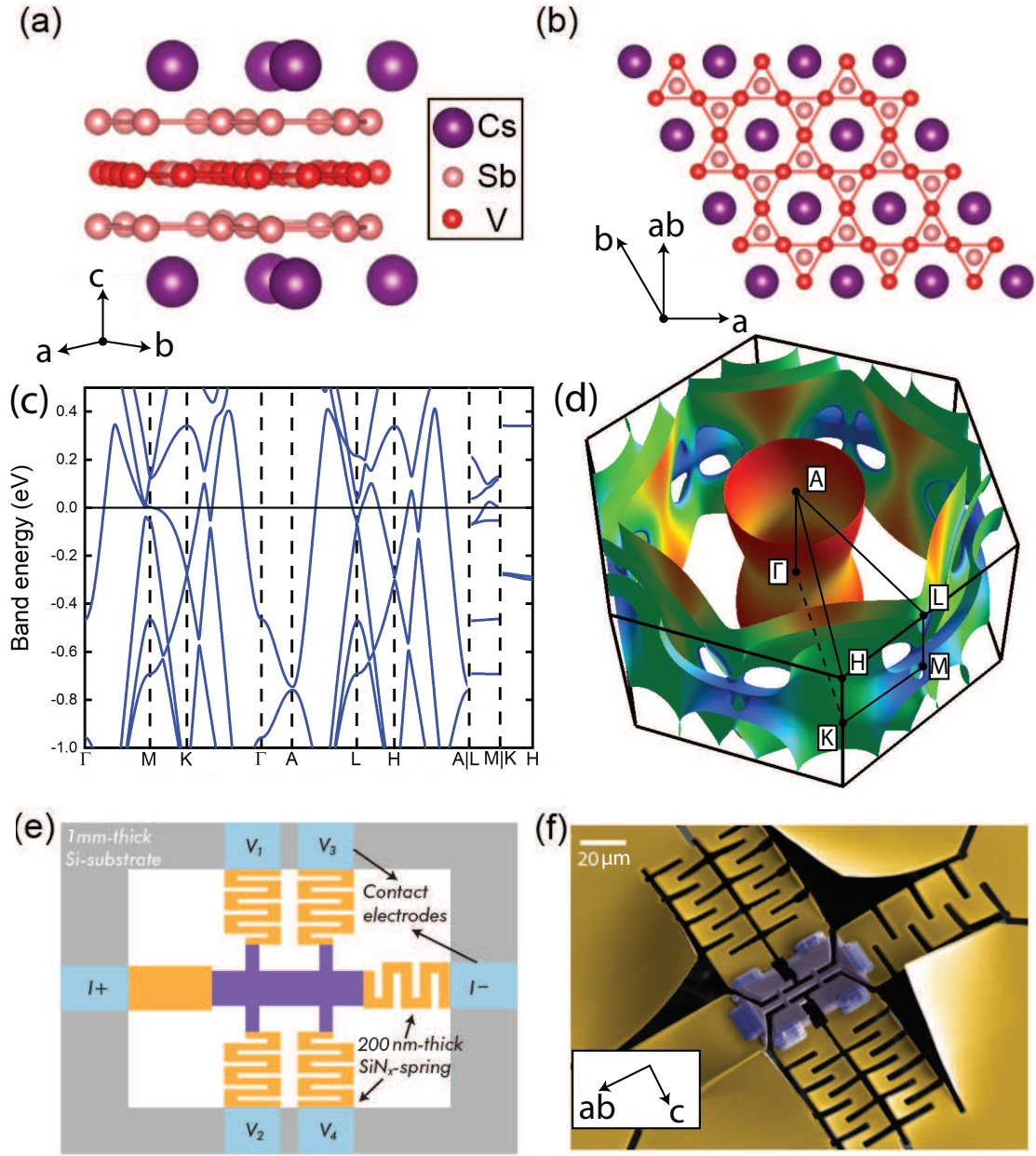}
		\caption{Side(a) and top(b) view of crystal structure of CsV$_3$Sb$_5$. It consists with layers of Cs, V and Sb atoms which form the Kagome lattice plane. (c) Electronic  structure of CsV$_3$Sb$_5$ calculated by density-functional theory. There exist multiple electronic bands across the Fermi level. (d) Fermi surfaces in the Brillouin zone, including several types of Fermi surfaces and their symmetric copies. (e) Illustration of membrane-mount CsV$_3$Sb$_5$ device. The sample is suspended with soft SiN$_x$ springs, which ensures the minimum strain effect. (f) Scanning-electron-microscope (SEM) image of the Membrane device. The 250 by 250 $\mu$m membrane window was patterned into several branches of springs with FIB.}
	\label{Crystal structure}
\end{figure}

\begin{figure}
	\centering
	\includegraphics[width = 0.85\linewidth]{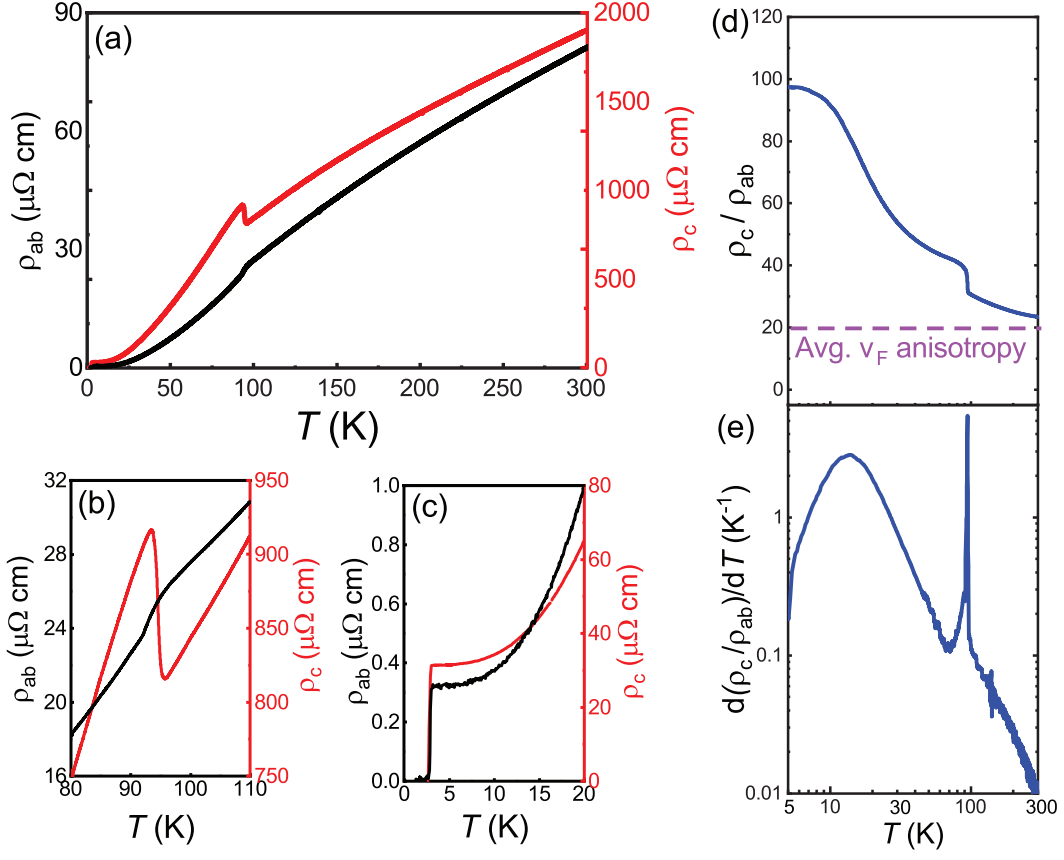}
		\caption{(a) Temperature dependence of the out-of-plane($\rho_c$) and in-plane resistivity($\rho_{ab}$). (b) An abrupt jump occurs in the temperature dependence of $\rho_c$, which corresponds to the charge-density-wave transition temperature $T_{CDW}$. Consistently $\rho_{ab}$ also displays a weak discontinuity. (c) Enlarged view of (a) within the low temperature range which demonstrates a clear superconducting transition at $T_c$ = 2.8 K. (d) Temperature dependence of resistivity anisotropy ($\rho_{c}/\rho_{ab}$). The dashed lines stand for the averaged Fermi velocity anisotropy for all Fermi surfaces. (e) First-order derivative of temperature dependence of resistivity anisotropy [d($\rho_{c}/\rho_{ab}$)/d$T$]. At the charge-density-wave transition temperature ($T_{CDW}$) a sharp peak can be observed, as well as a broad hump at $T$" $\approx$ 14 K.}
	\label{RT}
\end{figure}
\clearpage
\begin{figure}
	\centering
	\includegraphics[width=1\columnwidth]{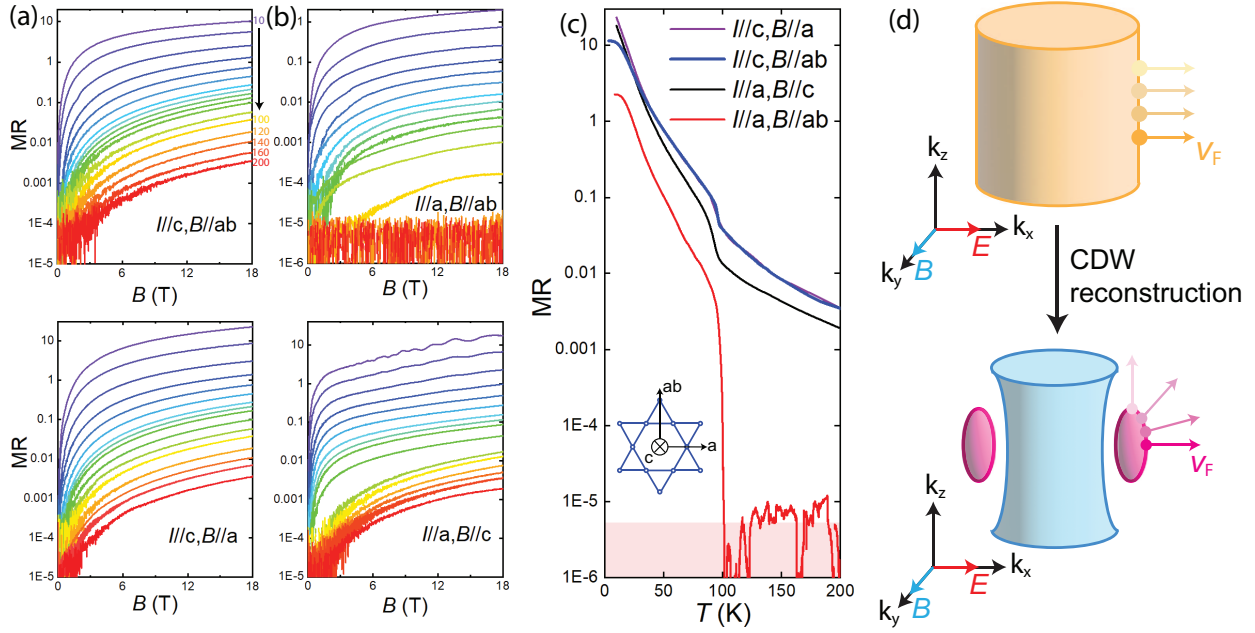}
		\caption{Field dependence of out-of-plane (a) and in-plane (b) magnetoresistance ratio [MR = ($\rho$ (18~T)-$\rho$ (0~T))/$\rho$ (0~T)] with field applied along a and c(ab)-axis. (c) Temperature dependence of magnetoresistance ratio at $B$ = 18~T with current along ab-axis and field along a-axis which are defined in the inset. (d) Illustration of Fermi surface reconstruction due to charge ordering. The abrupt appearance of magnetoresistance below $T_{CDW}$ is a direct consequence of the 3D pockets appears only after the CDW reconstruction. }
	\label{MRAniso}
\end{figure}
\clearpage

\begin{figure}
	\centering
	\includegraphics[width = 0.9\linewidth]{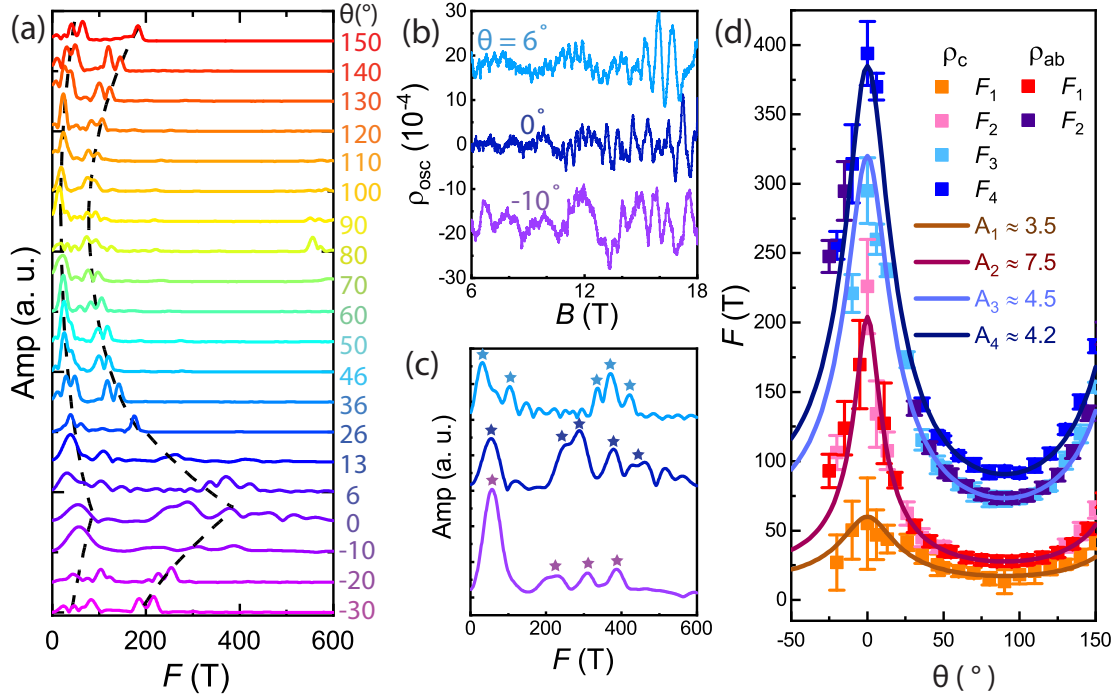}
		\caption{(a) Fast-Fourier-transformation(FFT) spectrum with different field directions for Shubnikov-de-Hass(SdH) oscillations measured with current applied along c-axis. Here $\theta$ stands for the angle between a-axis and field direction. (b) SdH oscillations ($\rho_{osc}$) at $\theta$ = 6, 0 and -10$^\circ$. {Here $\rho_{osc}=\Delta \rho / \rho_{BG}$, with $\Delta \rho$ the oscillatory part of the magneto-resistivity, and $\rho_{BG}$ a polynomial fit to the magneto-resistivity background.} (c) Corresponding FFT spectrum with the star symbols indicate the identified peaks. (d) Angular dependence of SdH oscillation frequencies. The error bar is defined as full width at half maximum of the peaks in the FFT spectrum.}
	\label{SdH}
\end{figure}

\end{document}


\renewcommand{\theequation}{S\arabic{equation}}
	\newcommand{\beginsupplement}{%
		\setcounter{table}{0}
		\renewcommand{\thetable}{S\arabic{table}}%
		\setcounter{figure}{0}
		\renewcommand{\thefigure}{S\arabic{figure}}%
	}
\title{Supplementary materials for "3D Fermi surfaces from charge order in layered CsV$_3$Sb$_5$"}

\author{Xiangwei Huang${}^{}$}\affiliation{Laboratory of Quantum Materials (QMAT), Institute of Materials (IMX),\'{E}cole Polytechnique F\'{e}d\'{e}rale de Lausanne (EPFL), CH-1015 Lausanne, Switzerland}
\author{Chunyu Guo${}^{\ast}$}\affiliation{Laboratory of Quantum Materials (QMAT), Institute of Materials (IMX),\'{E}cole Polytechnique F\'{e}d\'{e}rale de Lausanne (EPFL), CH-1015 Lausanne, Switzerland}
\affiliation{Max Planck Institute for the Structure and Dynamics of Matter, 22761 Hamburg, Germany}
\author{Carsten Putzke${}^{}$}\affiliation{Laboratory of Quantum Materials (QMAT), Institute of Materials (IMX),\'{E}cole Polytechnique F\'{e}d\'{e}rale de Lausanne (EPFL), CH-1015 Lausanne, Switzerland}
\affiliation{Max Planck Institute for the Structure and Dynamics of Matter, 22761 Hamburg, Germany}
\author{Martin Gutierrez}
\affiliation{Department of Physics, University of the Basque Country (UPV/EHU), Apartado 644, 48080 Bilbao, Spain}
\affiliation{Centro de Física de Materiales (CSIC-UPV/EHU), 20018 Donostia-San Sebastian, Spain}
\author{Yan Sun}\affiliation{Max Planck Institute for Chemical Physics of Solids, 01187 Dresden, Germany}
\author{Maia G. Vergniory}\affiliation{Max Planck Institute for Chemical Physics of Solids, 01187 Dresden, Germany}
\affiliation{Donostia International Physics Center, 20018 Donostia-San Sebastian, Spain}
\author{Ion Errea}\affiliation{Donostia International Physics Center, 20018 Donostia-San Sebastian, Spain}
\affiliation{Centro de Física de Materiales (CSIC-UPV/EHU), 20018 Donostia-San Sebastian, Spain}
\affiliation{Fisika Aplikatua Saila, Gipuzkoako Ingeniaritza Eskola, University of the Basque Country (UPV/EHU), 20018 Donostia-San Sebastian, Spain}

\author{Dong Chen}\affiliation{Max Planck Institute for Chemical Physics of Solids, 01187 Dresden, Germany}
\author{Claudia Felser}\affiliation{Max Planck Institute for Chemical Physics of Solids, 01187 Dresden, Germany}
\author{Philip J. W. Moll${}^{\dagger}$}\affiliation{Laboratory of Quantum Materials (QMAT), Institute of Materials (IMX),\'{E}cole Polytechnique F\'{e}d\'{e}rale de Lausanne (EPFL), CH-1015 Lausanne, Switzerland}
\affiliation{Max Planck Institute for the Structure and Dynamics of Matter, 22761 Hamburg, Germany}
\date{\today}

\maketitle
\beginsupplement

\section{Fabrication process of membrane-based microstructures}
The fabrication of the suspended microstructure starts by cutting a cross-sectional slice of the platelet in the $ac$-plane by FIB milling. A Xe-ion beam is used to avoid ion implantation, and a low-voltage polish step at 5~kV reduces the amorphization layer into the nm-range\cite{kelley2013xe+}. This slab is then transferred to a commercial SiN$_x$ transmission electron microscopy window (SPI). The 500$\times$500 $\mu$m wide window in a Si frame is spanned by a 200~nm thin membrane. After Au-evaporation, the sample is further patterned into a symmetric transport geometry and the remaining window is cut into a meandering spring, to further soften the mount guided by finite element simulations. We have fabricated both in-plane and out-of-plane aligned microbars that are maximally decoupled from its mechanical support[Fig. 1(f)]. They feature a minimal residual pressure that allows to probe reliably transport bars of materials which easily cleave ($\approx$ 9.8~bar, see SEC. II).

\section{Estimation of strain due to thermal contraction for MEM-based microstructure}
We now estimate the displacement applied to the sample from the thermal contraction of the SiN$_{x}$ spring and the outer Si frame when cooling the device to cryogenic temperatures. The thermal contraction coefficient of SiN$_{x}$($\upvarepsilon_{SiN}$) and Si($\upvarepsilon_{Si}$) upon cooling from 300~K to 4~K are 0.0342\% and 0.0208\% respectively. Assuming the sample itself has a typical thermal contraction coefficient $\upvarepsilon_{Samp}$ $\approx$ 0.1\%, one can easily calculate the total thermal contraction of the SiN$_x$ spring:

\begin{align}
dL_{Samp}=L_{Samp} \times \upvarepsilon_{Samp} = 80~nm&
\\dL_{SiN}=L_{SiN} \times \upvarepsilon_{SiN} = 70~nm&
\\dL_{Si}=L_{Si} \times \upvarepsilon_{Si} = 60~nm&
\\dL_{total}=dL_{Samp}+dL_{SiN}-dL_{Si} = 90~nm&
 \end{align}
 
Since the spring constant of the the SiN$_x$ spring is determined as 110 N/m by Comsol multiphysics simulation, the pressure due to thermal contraction can be calculated as: 

\begin{equation}
P_{samp}=k \cdot dL / A = 9.8~bar
 \end{equation}
where A stands for the cross section of the spring.

\section{Details of band structure calculations}
First-principles density functional theory (DFT) calculations were performed using the Quantum Espresso package (QE) \cite{QE-2017}. We used the generalized gradient approximation with the Perdew-Burke-Ernzerhof parameterization \cite{perdew1996generalized} together with projector-augmented wave pseudopotentials generated by Dal Corso \cite{DALCORSO2014337}. In all cases, we used an energy cutoff of 90 Ry with a Methfessel-Paxton smearing \cite{MP-smearing} of 0.02 Ry. The structural relaxation was done with a 18x18x2 grid neglecting spin-orbit coupling (SOC) and was stopped when pressures are below 0.01 kBar. The bands were computed adding SOC in top of the previous parameters. For the calculation of the Fermi surface we performed a wannierization of the system using Wannier90 \cite{mostofi_2014} and then computed the Fermi surface in a 200x200x100 grid using the obtained tight binding model.

\clearpage